\documentstyle[prl,aps,multicol]{revtex}
\input{epsf}
\def\kbar{\protect\@kbar}
\def\@kbar{%
\relax \bgroup
\def\@tempa{\hbox{\raise.73\ht0
\hbox to0pt{\kern.25\wd0\vrule width.5\wd0
height.1pt depth.1pt\hss}\box0}}%
\mathchoice{\setbox0\hbox{$\displaystyle k$}\@tempa}%
{\setbox0\hbox{$\textstyle k$}\@tempa}%
{\setbox0\hbox{$\scriptstyle k$}\@tempa}%
{\setbox0\hbox{$\scriptscriptstyle k$}\@tempa}%
\egroup}
\newcommand{\Uop}{{\hat U}_\tau}
\newcommand{\Dp}{\partial}

\begin{document}

\draft
\title{Effects of a nonlinear perturbation on dynamical tunneling in cold 
atoms.}

\author{Roberto Artuso $^{1,2,3}$, Laura Rebuzzini$^{1}$}

\address{$^1$ Center for Nonlinear and Complex Systems, 
Universit\`a dell'Insubria a Como, Via Valleggio 11, 22100 Como, Italy}
\address{$^2$  Istituto Nazionale per la Fisica della Materia, 
Unit\`a di Como, Via Valleggio 11, 22100 Como, Italy}
\address{$^3$ Istituto Nazionale di Fisica Nucleare, Sezione di Milano, 
Via Celoria 16, 20133 Milano, Italy}


\maketitle

\begin{abstract}
{We perform a numerical analysis of the effects of a nonlinear 
perturbation on the quantum 
dynamics of two models describing non-interacting cold atoms 
in a standing wave 
of light with a periodical modulated amplitude $A(t)$. One model is the driven 
pendulum, considered in ref.\cite{raiz1}, and the other is a variant of the 
well-known Kicked Rotator Model.
In absence of the nonlinear perturbation, the system is invariant under some
discrete symmetries and quantum dynamical 
tunnelling between symmetric classical islands is found. 
The presence of nonlinearity 
destroys tunnelling, breaking the symmetries of the system. Finally, 
further consequences of nonlinearity in the kicked rotator case are 
considered.}

\end{abstract}                      



\begin{multicols}{2}
\narrowtext
 
Tunneling is one of the most typical features of quantum mechanics, 
concerning oscillations between states that cannot be connected in the 
classical Hamiltonian dynamics by real trajectories. The original formulation 
of the problem involved states separated by potential barriers: the relevant 
features are semiclassically explained in terms of complex solutions of 
Hamilton equations for one-dimensional systems \cite{BB}, while a proper 
treatment in higher dimensions is considerably subtler even when integrability
is preserved \cite{WIL}. Recently a novel kind of tunneling, involving 
transitions between classically separated regions in the presence of a non 
trivial structure of the phase space, has attracted considerable attention, 
both theoretically \cite{DH,TU,MM,MD}, and experimentally \cite{DG,raiz1,HH}.
In particular a large fraction of these papers focus on physical settings
realized with cold atoms in optical potentials. The recent 
widespread interest and experimental activity in Bose-Einstein condensation 
\cite{BEC} suggest to check whether the presence of Gross-Pitaevskii 
nonlinearities \cite{GP} deeply influences the characteristic features 
of dynamical tunneling. Such a question was already raised and analyzed 
in the framework of the kicked oscillator \cite{KO}, where it was 
observed how the nonlinear terms typically destroys quantum effects induced 
by symmetry \cite{NKO,roblau}. 
The paper is organized as follows: we firstly give a few details 
and fix notations for the class of models we are going to consider, then 
analyze two cases: a driven pendulum and the kicked rotator;
 in the last section we briefly reconsider pioneering work made on the 
nonlinear kicked rotator \cite{cas,dima} and supplement it with novel results
for the quantum resonant case.

\section {General setting.}

We will consider models described by the following hamiltonian
\begin{equation}
\label{ham}
H=H_O + V(\vartheta ,t)= \frac{p^2}{2} +A(t) cos\vartheta,
\end{equation}
where $p$ and $\vartheta$ are canonical momentum and position coordinates 
respectively, expressed in scaled dimensionless units, and $A(t)$ 
is a periodic function of time. 
Under appropriate choices of the function $A(t)$, the quantum version of 
this model describes an ensemble of non-interacting cold atoms 
in presence of a standing wave with a periodically
 modulated amplitude $A(t)$.
The connection between 
the scaled variables $\vartheta,p$ and the physical ones 
$\vartheta',p'$
 is
$\vartheta =2k_L\vartheta^{'}$ and $p={\kbar} 
p^{'}/2k_L\hbar$ 
( where $k_L$ is half of 
the wave vector of the standing wave and $\kbar$ is the effective Planck
 constant $[\vartheta,p]=-i\kbar$)\cite{raiz2}.
\begin{figure}
\centerline{\epsfxsize=7cm \epsfbox{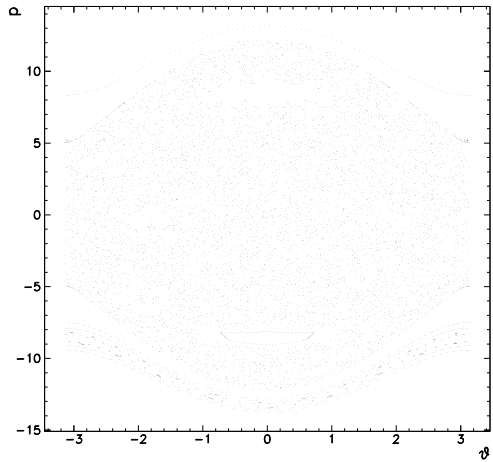}}
\caption{\small 
Poincar\'e section of an ensemble of classical particles 
for the driven pendulum. The 
initial ensemble is  
uniformly distributed in $|p|<10$, $\theta\in [-\pi,\pi[$; 
the values of $p$ and $\theta$ are taken at 
intervals multiples of the period $\tau$. }
\label{fig1}
\end{figure}
We consider two possible choices for the periodic function $A(t)$:
\begin{itemize}
\item {The first model is the driven pendulum. The amplitude $A(t)$ is that 
reported in
 \cite{raiz1}, i.e. $-2\alpha \cos^2(\pi 
t)$; the period $\tau$ of time modulation is 1 and the parameter $\alpha$ 
depends 
on some physical quantities of the system, kept fixed 
in experiments, such as the ac Stark shift amplitude, the electric field 
strength and the dipole momentum of the atom.}

\item {The second model is a variant of the 
well-known Kicked Rotator model \cite{kr1,raiz2}, where 
the amplitude $A(t)$ is a sum of $\delta$ 
functions of period $\tau$,
$k \cos\vartheta
 \sum _{n=0}^{n=+\infty}\delta (t-n\tau )$. The parameter $k$ 
measures the kick strength.} 
\end{itemize}

\begin{figure}
\centerline{\epsfxsize=7cm \epsfbox{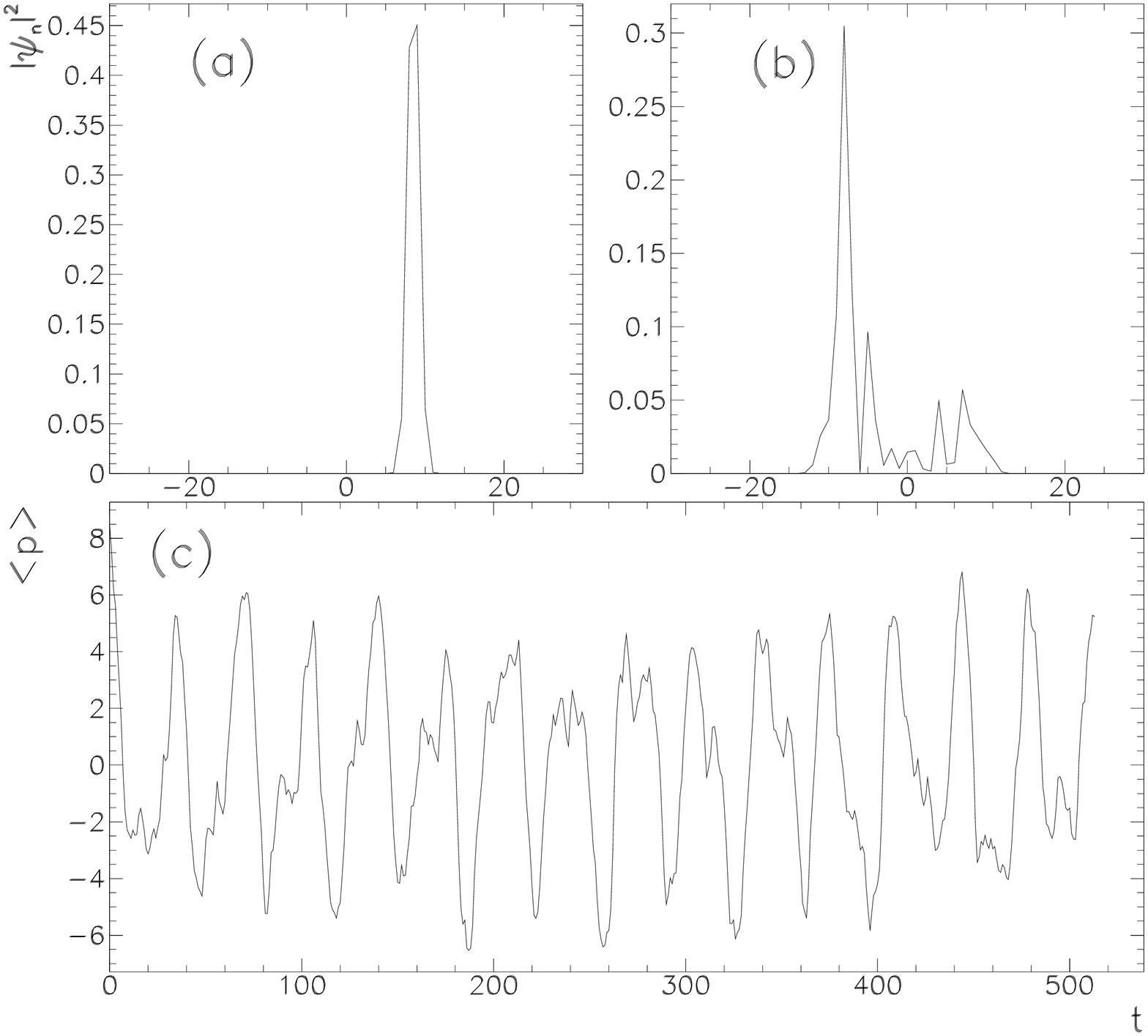}}
\caption{\small (a) The momentum 
distributions at times $t=0$ and (b) $t=56\tau$ 
for one rotor with zero quasi-momentum. 
 The values of the parameters 
are $\alpha =10.0$, $N=128$, $L=500$. (c) The correspondent 
tunnelling oscillations of the first moment. 
Time is measured in number of periods.}
\label{fig2}
\end{figure}

Owing to the time dependence of the amplitude $A(t)$, in the Kicked Rotator 
time can be treated as a discrete variable, measured in intervals of 
the period $\tau$, and the evolution is discrete;
 in the driven pendulum instead, time is a continuous variable. 

The two models present some analogies.
In both models the potential $V(\vartheta,t)$ is periodic both 
in time and space 
with period $\tau$ and $2\pi$ respectively. 
Moreover, the two models share some symmetries:
time-reversal ($(\vartheta , p,t)\to  (\vartheta , -p,-t)$) and  
parity ($(\vartheta , p,t)\to  (-\vartheta , -p,-t)$).

The quantum Hamiltonian operator of the unperturbed system is obtained from 
the Hamiltonian 
(\ref{ham}) replacing the classical canonical variables with the correspondent 
operators $\hat \vartheta$ and $\hat p=-i\hbar\Dp /\Dp \hat\vartheta$. 

Space and time periodicity are conveniently dealt with by using Bloch 
Floquet Theory. As the potential commutes with spatial translation of 
$2\pi$, the quantized momentum gets eigenvalues on a discrete lattice 
$p=\hbar n +\beta$, $n\in {\bf Z}$ and where $\beta$, called the 
quasimomentum, is the analogous of the Bloch phase in solid state physics 
\cite{AM}. In the present paper we choose $\hbar =1$ (see the former 
discussion about choice of units), so that the quasimomentum takes values 
in the interval $[0,1)$ (the analogous of the first Brillouin zone).

Quasi-momentum is preserved during quantum 
evolution and parameterizes
a single quantum rotor, fixing the origin of the discrete lattice in momentum 
space.
The evolution of an  
ensemble of non interacting atoms can be modelled by a superposition of the 
evolutions of 
independent rotors.

Because of the invariance of the quantum evolution operator under time 
translation 
by an interval $\tau$ ($U(t+\tau)=U(t)$), an evolution operator $U_{\tau}$
over one period can be defined, called Floquet 
operator.
 The eigenvalues of Floquet 
operator are $e^{-i\frac {\epsilon\tau}{\hbar}}$, where the real 
quantities $\epsilon$, independent from $\tau$, are called 
quasi-energies. The quasi-energy spectrum is invariant under translation 
over $2\pi\hbar /\tau$. Quasi-energy plays the same role of energy 
in systems with continuous time variable.

Concerning the discrete symmetries we mentioned, 
note that changing the sign of $p$ means 
to change the sign of the integer part of $p$ ($[p]=n\to -n$) and to replace 
the fractional part $\{p\}$ of $p$ by $1-\{p\}$ ($\beta\to 1-\beta$). 
Therefore 
only rotors with $\beta =0$ or $\beta =\frac 12$ are invariant with 
respect to these 
discrete symmetries; rotors with 
$\beta \neq 0,\frac 12$ mix each others: a rotor labelled by $\beta$ is 
mapped to another rotor labelled by $\beta ' =1-\beta$.

Now we discuss some feature of quantum dynamical tunneling for the 
two cases we selected, also considering the 
effect of Gross-Pitaevskii nonlinearities.

\begin{figure}
\centerline{\epsfxsize=7cm \epsfbox{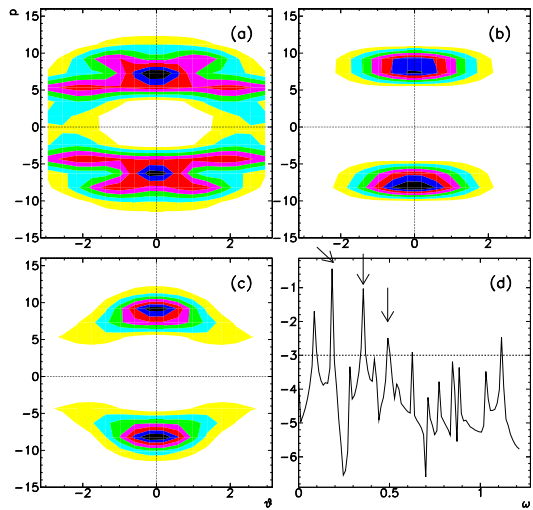}}
\caption{\small (a),(b),(c) Three Floquet eigenstates involved in the 
dynamics of the system; 
(d) the Fourier spectrum of $\langle p(t)\rangle $. The dashed horizontal line 
determines the limit of the dominant frequencies: the five 
higher peaks emerge.}
\label{fig3}
\end{figure}
 
\section { The driven pendulum.}

Firstly we analyze the unperturbed system (\ref{ham})
without the nonlinear term, which is the theoretical setting for 
experimental data reported in ref.\cite{raiz1}.

Note that, in the experimental results reported in \cite{raiz1},  
the momentum 
is measured in units of $2k_L \hbar$, 
which correspond to the scaled momentum $p$ divided by $\kbar\approx 2.08$.

The classical trajectories are obtained by a numerical integration of the 
Hamilton equations with fourth-fifth order Runge-Kutta method. 

In Fig.1 the Poincar\'e 
 surface of section for positions $\vartheta$ and momenta 
$p$ of an ensemble of classical particles is shown for the value $\alpha 
=10.5$:
the time evolution 
of 1000 initial conditions, uniformly distributed in the square $|p|<10$, 
$\vartheta\in [-\pi,\pi[$, 
is recorded at 100 subsequent modulation periods. The symmetries of the 
system 
under the transformations 
$\vartheta \to -\vartheta$ and $p\to -p$ are clear.
A pair of stable fixed points of period two $(\vartheta_0 =0; p_0=\pm 8.65)$ 
are present. The stability islands, formed by regular trajectories surrounding 
the fixed points, are time-reversed images of each other.

\begin{figure}
\centerline{\epsfxsize=7cm \epsfbox{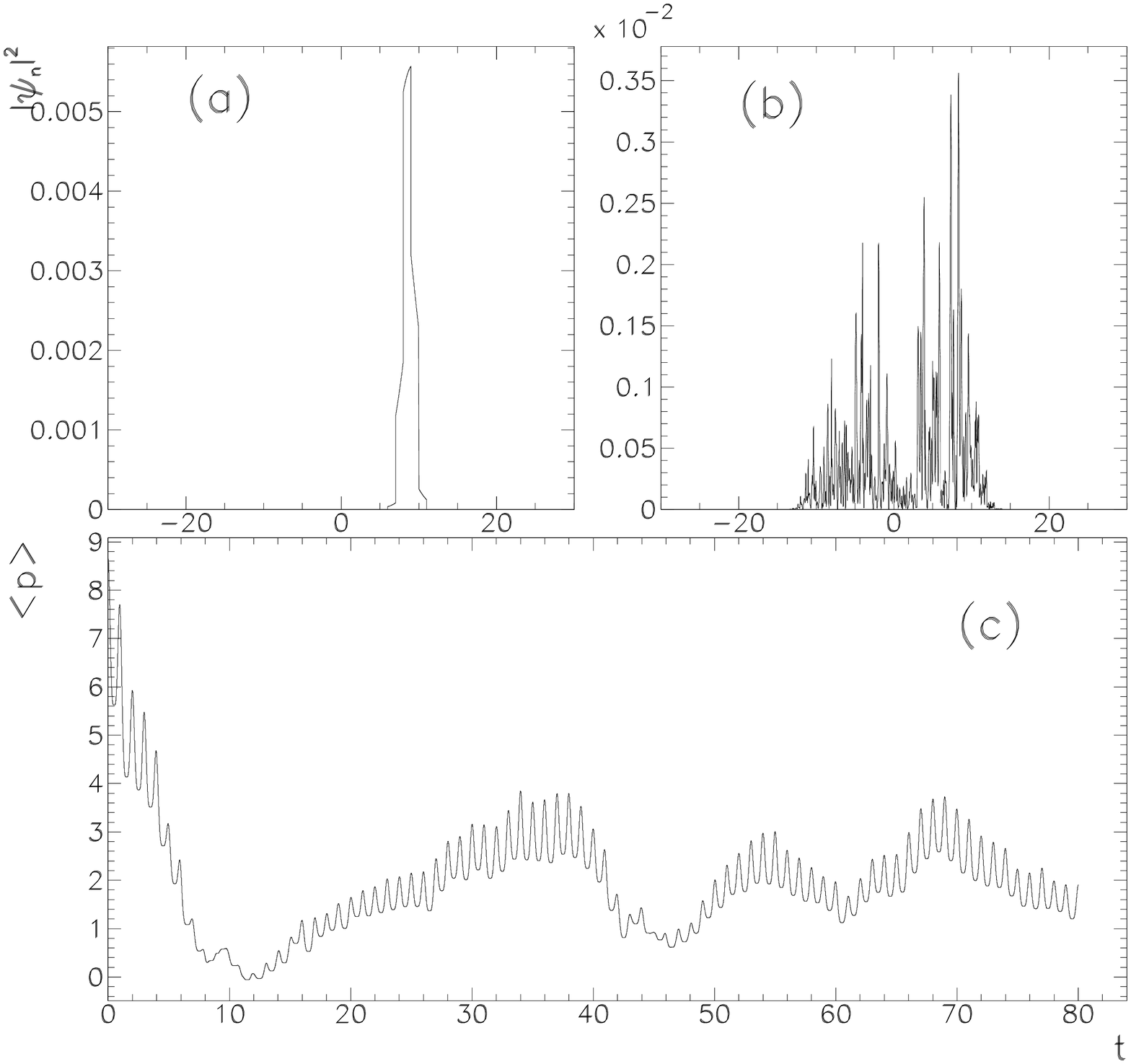}}
\caption{\small The momentum 
distributions at times $t=0$ and (b) $t=10\tau$ in a system of 101 
$\beta$-rotors, with $\beta$ equally spaced in the 
interval $0.4\leq\beta \leq0.6$. The values of the parameters 
are $\alpha =10.5$, $N=128$, $L=500$. (c) The correspondent 
tunnelling oscillations. Time is measured in number of periods.}
\label{fig4}
\end{figure}

Since the Hamiltonian operator is non autonomous, the 
exact Floquet operator of a quantum rotor, with fixed quasi-momentum 
$\beta$, is given by a Dyson expansion.
In the numerical simulation
 we use the lowest order split method \cite{split}. 
The Floquet operator is approximated by an ordered 
product of evolution operators on a small intervals of time $\Delta t =\tau 
/L$, with $L$ integer:
\begin{equation}
\label{Fop}
\begin{array}{l}
 \Uop (\beta )= T \exp ^{-i\int _{0} ^{\tau } 
\left( \hat H_0 +\hat V(\hat\vartheta ; t)\right)} \\
 \approx \prod _{k=1}^{L} \exp ^{-i\hat H_0 \frac {\Delta t}{2}}
\exp ^{-i\hat V(\hat\vartheta; t) \Delta t} 
\exp ^{-i\hat H_0 \frac {\Delta t}{2}}\\
 = \prod _{k=1}^{L} \exp ^{-i\frac {\tau}{4L}(\hat n +\beta )^2 }
\exp ^{-i\frac {\tau}{L}2\alpha \cos ^2\left(\pi \frac {\tau}L k\right)
\cos\hat\vartheta }\\
\ \ \ \ \ \exp ^{-i\frac {\tau}{4L}(\hat n +\beta )^2 }.\\
\end{array}
\end{equation}

We use a finite base of dimension $N$: 
the discrete momentum eigenvalues belong to the finite lattice $p=(m-\frac 
N2)+\beta$ and the continuous angle variable is approximated by 
$\vartheta =\frac {2\pi}N(m-
1)$ with $m\in {\bf Z}, 1\leq m\leq N$. 

We start by considering the case of a simple rotor, with quasimomentum 
$\beta =0$.

In fig. 2 the 
evolution of the first moment and the correspondent 
momentum distribution are shown
for $\alpha =10$.
The initial state is a coherent state 
centered in one of the classical symmetric 
islands:
\begin{equation}
\label{inst}
f(p)=C\exp^{-ip\vartheta_0}
\exp^{-\delta^2 (p-p_0)^2},
\end{equation}
where the constant $\delta$ is $\sqrt {1/2}$ and $C$ 
a normalization factor. In the simulations we use $\hbar =1$, 
$\vartheta_0 =0$ and $p_0 =8.525$.
\begin{figure}
\centerline{\epsfxsize=7cm \epsfbox{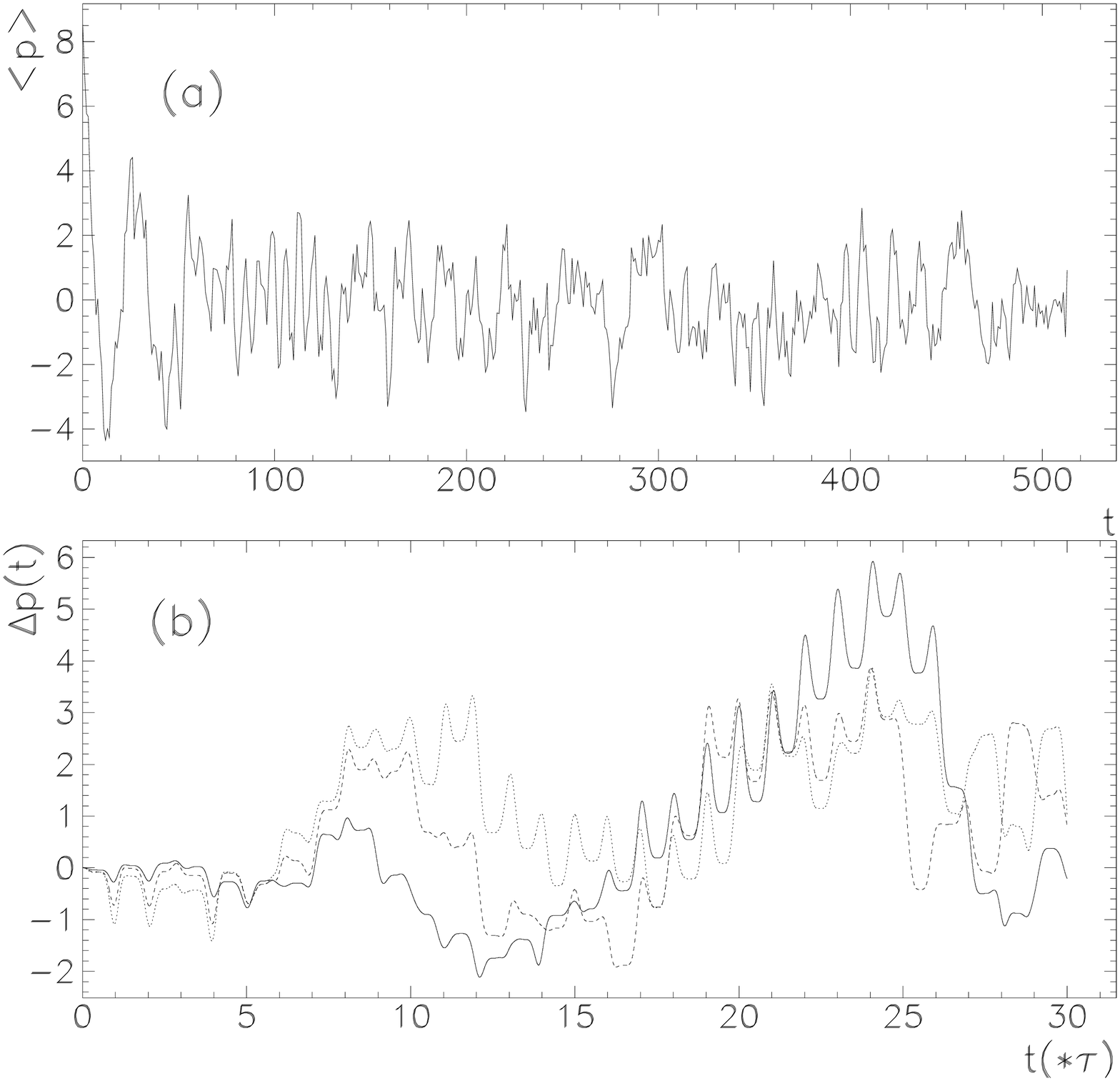}}
\caption{\small (a)$\langle p(t)\rangle$ for $u=5,\alpha =10,N=128$;
(b)  $\langle p(t)\rangle_u -\langle p(t)\rangle_0$ versus t, 
for $u=3$(solid line),$7$(dashed line),$10$(dotted line).Time is measured in 
number of periods. }
\label{fig5}
\end{figure}

It can be seen that the 
maximum of the 
distribution probability oscillates periodically 
between the two symmetric values $+p_0$ and $-p_0$. 
In fig.2(b) 
the  probability density is shown at time $t=56\tau$ corresponding 
to a negative value of $\langle p\rangle$: its maximum is peaked at $n=-9$.
As we already remarked, for $\beta =0$ the quantum 
system is invariant with respect to parity $\hat P$ 
and time inversion $\hat T$ and dynamical tunnelling between the symmetric 
classical stable regions is present, marked by quasi-periodical oscillations 
of the first moment between symmetric 
positive and negative values (approximately $\langle p\rangle 
\approx \pm 6$). For values of 
quasimomenta $\beta$ different 
from $0$ or $1/2$, exact quantum symmetries are broken and periodical
 oscillations are damped and then suppressed. The damping of oscillations is 
faster for values 
of $\beta$ far from 0 and 1/2.

The time evolution of the first moment can be expanded in terms of 
Floquet eigenstates as:
\begin{equation}
\label{pmed}
\langle p(t)\rangle =\sum_{j,l}c_j (0)c^*_l(0) \exp^{-\frac i\hbar t(\phi_j 
-\phi_l)}\sum _n n\chi_j(n)\chi^*_l (n)
\end{equation}
where the Floquet eigenfunctions $\chi_{j} (n)$ are expressed in momentum 
representation, the phases $\phi _{j,l}$ are the correspondent eigenvalues 
(connected to quasi-energies by $\phi =\tau \epsilon$)
and the squared 
modulus  of the coefficients $c_j (0)=\langle \chi_j |\psi (0)\rangle $ gives 
the overlap of the eigenfunction $|\chi _j\rangle$ 
with the initial state.
As it can be seen from (\ref{pmed}), 
the Fourier frequencies $\omega _{j,l}$, in 
which the periodic motion can be decomposed, are 
separations  
between two Floquet eigenvalues ($\omega _{j,l} = \phi _j -\phi _i$). The 
dominant frequencies $\omega _{j,l}$ correspond to 
Floquet eigenstates that more contribute to the dynamics, having
 an high overlap 
with the initial state.
\begin{figure}
\centerline{\epsfxsize=7cm \epsfbox{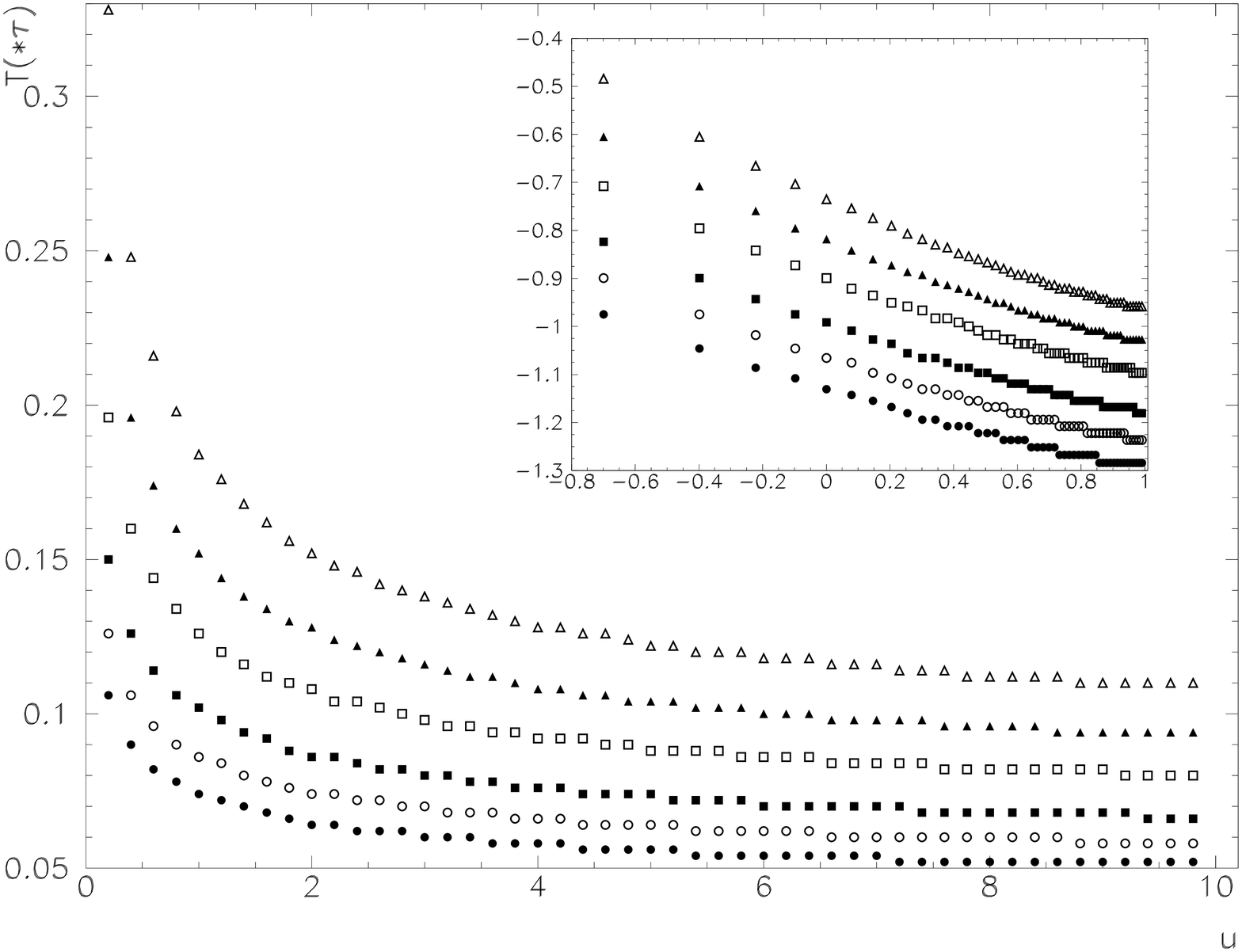}}
\caption{\small The separation time $T$ between $P_u$ and $P_0$,  
when $|P_u (T)-P_{u=0}(T)|$ exceeds the value $\Delta$ versus $u$. 
Starting from above, the six curves 
correspond to $\Delta =0.2, 0.1, 0.05, 0.02, 0.01 ,0.005$.  
In the inset a log-log plot is shown.}
\label{fig6}
\end{figure}

The Floquet eigenfunctions 
and eigenphases are evaluated by a numerical diagonalization of the operator.

By using a finite basis (the momentum is discretized using $N$ points), the 
Floquet operator is reduced to a finite matrix $N\times N$ and it is 
calculated as follow.
The columns of the Floquet operator in the momentum representation 
are obtained by evolving over one period $\tau$ 
the momentum eigenstates. 
Owing to the finiteness 
of the basis, the evolution of 
the momentum eigenstates initially 
localized on the edges of the basis is affected by errors.
To overcome this problem, we calculate the Floquet 
operator using a basis of double dimension  
$\tilde N =2N$, so that
the variable $p$ takes discrete values in the interval 
$p\in [-N+1, N ]$. 
Then we extract from the matrix $\tilde N\times\tilde N$ 
a nonunitary submatrix 
of dimension 
$N$ in which $p$ varies in the range $[-\frac N2 +1+ ,\frac N2 ]$.
The disadvantage of using a nonunitary matrix is to find some 
eigenvalues inside 
the unitary circle.
Nevertheless, the number of nonunitary eigenphases and the errors 
can be reduced by increasing the dimensions of the bases $N$ and 
$\tilde N$.

The pairs of eigenfunctions 
$(\chi _j ,\chi _i )$ that dominate 
the dynamics of the system are selected by three conditions \cite{floq}: a) 
the maximum presence probability $\bar p$ 
inside the region where the classical 
stable islands lie (i.e. $|n|<10$); b) the maximum overlapping probability of 
each 
eigenfunction with the initial state, $P_j =|\langle \chi _j |\psi (0)
\rangle |^2$; c) the 
maximum mutual overlapping probability between the two eigenfunctions,  
$P_j P_i$. 
In fig.3(a),(b),(c) three Floquet eigenstates 
verifying the conditions $\bar p _j > 0.5, P 
_j >0.07$ and $P _j P _i >0.0049$ are shown.

Fourier analysis of  $\langle p(t)\rangle $, with a resolution of 
$2\pi/512 \approx 0.0123$, reveals five dominant frequencies
($\omega _1 = 0.0859; \omega _2 =0.1841 ; 
\omega _3 = 0.3559
; \omega _4= 0.4909; \omega _5 = 1.1167$).
A part of the spectrum around zero is represented in fig.3 (d). The dominant 
frequencies are the peaks higher than the dashed horizontal line.

 The differences between the eigenphases of the eigenfunctions shown in fig.3 
 ($ \phi_a =4.4579, 
\phi_b =4.1035, \phi_c =3.9198$) 
correspond to the frequencies 
of the spectrum, marked by arrows: 
$\omega _2 = \phi_b -\phi_c$, $\omega _3 = 
\phi_a -\phi_b$, $\omega _4 = \phi_a -\phi_c$. 

Up to now we have considered quantum evolution corresponding to a fixed
quasimomentum (that furthermore respects quantum symmetries): we now consider
the evolution of a distribution of rotors with different quasimomenta $\beta$, 
each evolving with the operator (\ref{Fop})
 with a fixed quasi-momentum 
$\beta$\cite{fgr}: $\psi 
(p,0)=f(p)\delta _{{n},{[p]}}$. The initial momentum distribution $f(p)$ is a 
coherent state peaked in the center 
of one of the two classical islands
 of fig.1, as in (\ref{inst}).

During the evolution each $\beta$-rotor evolves 
independently and the mean value of the 
observables of the system is an average over different rotors. 
The time 
evolution of the momentum distribution   
and of the mean value of $\hat p$ has been calculated. The data for 
$\alpha =10.5$ are shown in fig.4. 
The spreading in 
quasi-momentum is $\Delta\beta
=0.2$ around the value $\beta=\frac 12$, which preserves the symmetries.
The 
oscillations of $\langle p(t)\rangle $ correspond to the
tunnelling of the quantum state between the two symmetric 
islands in the classical phase space. Note that, as in the experimental data 
\cite{raiz1}, during time evolution, 
the momentum distribution of the system 
 maintains its maximum localized in the 
starting island, in the region of positive momenta (see the momentum 
distribution in fig.4(b), corresponding to a minimum value of the average 
momentum).Therefore, in consequence of the average over different rotors, the 
oscillations 
of $\langle p(t)\rangle $ do not reach negative values, 
for which higher values of probability density in regions of 
negative momenta are needed.

Moreover the spreading in quasi-momentum can reproduce the 
decay of the oscillations in time. 
In accordance with \cite{MD}, we have also verified 
that the decay of oscillations is faster if the dispersion 
in $\beta$ is increased. 

As it can be seen from short-period oscillations in fig.4(c), the behaviour of 
$\langle p(t)\rangle $ is a superposition of 
tunnelling oscillations 
with slightly different frequencies, because 
the spreading in quasimomenta corresponds to a spreading of 
the dominant frequencies of the motion.

We now study the influence of a Gross-Pitaevskii nonlinearity in the 
dynamics: where Schr\"{o}dinger operator is modified by adding 
a nonlinear term dependent on the 
squared modulus of the wave function of the system, $u |\psi |^2$. 
The approximate expression of the evolution operator becomes:
\begin{equation}
\label{Fopnl}
\begin{array}{l}
\Uop  (\beta) \approx \prod _{k=1}^{L} \hat R\left(\frac \tau L\right) 
\exp ^{-i\frac {\tau}{L}2\alpha \cos ^2\left(\pi \frac {\tau}L 
k\right)\cos\vartheta }  \hat R\left(\frac \tau L\right) \\
\hat R\left(\frac \tau L\right) =
\exp ^{-i\frac {\tau}{4L}(\hat n +\beta )^2 }
\exp ^{-i\frac {\tau}{2L}u|\psi _n|^2}\\
\end{array}
\end{equation}
The multiplicative factor $\exp ^{-i\frac {\tau}{2L}u|\psi _n|^2}$ 
breaks the symmetries of the system.

In presence of the nonlinearity term, the tunnelling oscillations of the first 
moment loose periodicity in time and are progressively 
suppressed at long times (see fig.5(a)). 
In fig.5(b)  
the difference between the first moment of the unperturbed
 system and that of 
the perturbed one is shown for three values of the nonlinear parameter $u$: 
$u=3$ (solid line),
$u=7$ (dashed line), $u=10$ (dotted line).
\begin{figure}
\centerline{\epsfxsize=7cm \epsfbox{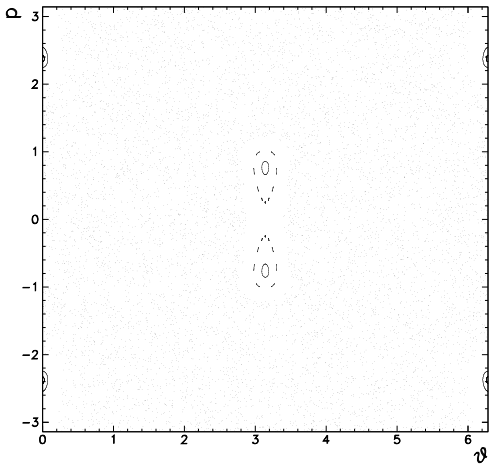}}
\caption{\small Classical phase space of the Standard Map (\ref{smapsim}) for 
$K=2.21$ with islands of accelerator modes. The phase space
is symmetric under space and time inversion. }
\label{fig7}
\end{figure}

We have also calculated the time-averaged first moment $P(t)=\frac 1t 
\sum _{t'=0}^{t-
1}\langle p(t')\rangle$ for fifty values of $u$ equally spaced in the 
interval $0< u< 10$ 
and we have compared the value of $P_u$ with $P_0$ in absence of nonlinearity. 

In fig.6 the time $T$ at which the quantity $|P_u (T)-P_{u=0}(T)|$ 
exceeds the fixed value $\Delta$ versus the nonlinear parameter $u$
is shown; 
each curve corresponds to a different value of the parameter $\Delta$ from 0.2 
to 0.005. 
The intervals of time 
$T$ decrease algebraically with respect to $u$ with the 
law $T=B/u^A$. 
In table 1 the power-law exponents $A$ 
and the constants $B$ are shown for different parameters $\Delta$.

\section { Kicked Rotator Model.}

The Hamiltonian of the classical system is:
\begin{equation}
\label{ham2}
H=\frac {p^2}2 +k\cos \vartheta \sum_{n=0}^{+\infty} \delta (t-n\tau)
\end{equation}

The correspondent classical map is the well-known Standard Map \cite{chir}, 
defined on the 
cylinder: 
\begin{equation}
\label{smap}
\left\{
\begin{array}{cc}
p_{n+1}=p_n +k\cos \vartheta_n   &  \nonumber\\
\vartheta_{n+1} =\vartheta_n + \tau p_{n+1} &   mod(2\pi)
\end{array}
\right.
\end{equation}
After the scaling of the variable $p$ ($p^{'}'=\tau p/2$) and the 
introduction of 
the parameter  $K=\tau k/2$, the map can be symmetrized and
 reduced on the 2-torus 
$[0,2\pi[\times [-\pi ,\pi[$:
\begin{equation}
\label{smapsim} 
\left\{
\begin{array}{cc}
\vartheta_{n+1}^{-} =\vartheta_n^{+} + p^{''}_{n} &   \nonumber\\
p^{''}_{n+1}=p'_n +K\cos \vartheta_{n+1}^{-}   & mod(2\pi)  \nonumber\\
\vartheta_{n+1}^{+} =\vartheta_{n+1}^{-} + p^{'}_{n+1} &   mod(2\pi)
\end{array}
\right.
\end{equation}
where the signs $+$ and $-$ refer to instants after and before the 
$n$-th and $n+1$-th kicks.

The classical system depends only on the parameter $K$. For $K\geq K_{c}
\approx 
0.97$ the system undergoes a transition to global chaos, the last K.A.M. curve 
breaks and unbounded diffusion in action space takes place. Nevertheless even 
for $K>>1$ some small stability islands survive in 
the classical phase space, correspondent to accelerator modes
\cite{chir,ll,Zas}.
\begin{figure}
\centerline{\epsfxsize=7cm \epsfbox{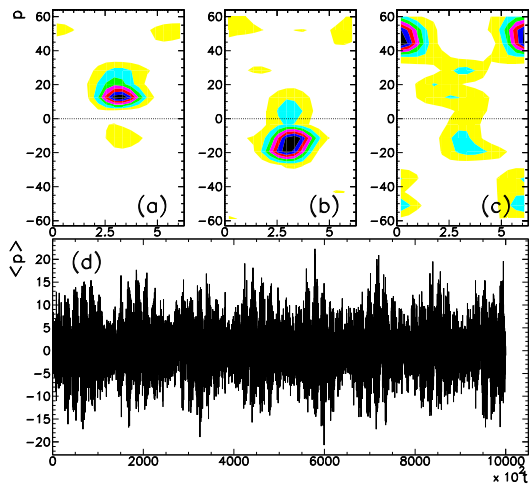}}
\caption{\small Husimi distribution of 
the state vector at times (a)t=800$\tau$,
 (b)t=900$\tau$, (c)t=8400$\tau$. (d) Tunnelling oscillations of 
$\langle p(t)\rangle$. Time is measured in number of periods.}
\label{fig8}
\end{figure}

In fig.7 the classical phase space of the map (\ref{smapsim}) for $K=2.21$ 
with the stability islands of accelerator modes is 
shown. 

\begin{figure}
\centerline{\epsfxsize=7cm \epsfbox{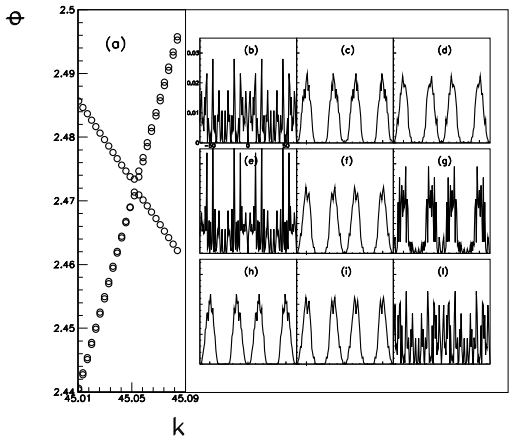}}
\caption{\small (a) A section
 of the eigenphases spectrum vs $k$ of the Floquet operator. 
Three eigenvalues for each values of $k$ are represented.
(b),(c),(d) The momentum distribution of the three correspondent
 eigenfunctions 
for $k=45.02$.
(e),(f),(g) The three correspondent eigenfunctions for $k=45.05$.
(h),(i),(l) The three correspondent eigenfunctions for $k=45.08$.}
\label{fig9}
\end{figure}
There are two periodic orbits of period two: one formed by the pair of fixed 
points $(\pi , \pm 0.7656)$ and the other by the pair 
$(0,\pm 2.3714$). The points of 
each periodic orbits are symmetric respect to $p=0$. They are surrounded by 
stability islands inside which the dynamics of the system is regular; each 
island is delimited by K.A.M. curves, which cannot 
be crossed.
During the evolution, ensembles of points initially located inside
 the islands centered in $(\pi ,\pm 0.7656)$ do not mix
with those  located inside the islands centered in $(0,\pm 2.3714)$.

The invariance of the system under space reflection and 
time inversion can be seen clearly from the structure of the phase space.

The quantum version of the system is a variant of the  
Kicked Rotator Model \cite{ccfi}, 
described by the Floquet operator:
\begin{equation}
\label{opkr}
\Uop = \hat R \hat K\hat R =
e^{-i\frac \tau 4 (\hat n+\beta )^2} e^{- i k\cos\vartheta}
e^{- i \frac \tau 4 (\hat n+\beta )^2}.
\end{equation}
(the kicked rotator corresponds to quasimomentum $\beta =0$).

For values of $\tau =4\pi M/N$ (with $M,N\in {\bf N}$) the kicked rotator  
undergoes a quantum resonance
\cite{izsh}, where the spectrum acquires a band 
structure, yielding ballistic transport (see also \cite{Fel}).

Having set $\hbar =1$, the parameters $\tau$ and $k$ are scaled by $\hbar$ 
with 
respect to the classical ones ($\tau =\tau_{cl} \hbar$ and $k=k_{cl}/\hbar$). 
Therefore the quantum parameter $\tau$ plays the role of $\hbar$; 
the classical 
limit of the system in obtained by the limits $\tau\to 0$, $k\to\infty$ and
$K=k\tau =const.$

For $\beta =0$ or $\beta =\frac 12$ the system is invariant respect to parity 
$\hat P$ and time 
inversion $\hat T$, as in the former case.
Therefore the Floquet eigenstates belong to invariant subspaces 
with respect to the 
discrete symmetries; for example they can be classified in two classes: odd or 
even with respect to parity \cite{CGG,LGrigW}.

In the following we fix the 
value of $\beta =0$ and analyze first the 
kicked rotator for a resonant value of $\tau$ and then 
for a generic value of $\tau$.

\subsection {Resonant case.}

Under the resonance condition $\tau =4\pi M/N$, the quantum system is 
periodic of period $N$, if $M$ or $N$ is even \cite{changshi}. 
Therefore the quantum system can be reduced on a torus 
and its Floquet operator 
becomes a unitary finite matrix of dimension $N\times N$. Exact 
eigenfunctions can be calculated.

The discrete momentum eigenvalues are $p^{'}=n=m-N/2$, 
with $m$ integer, varying 
in the interval $1\leq m\leq N$. To make a comparison between the 
quantum system 
and the classical one on the torus, the classical variable $p$ has to be 
rescaled by the factor $2/\tau$, i.e. $p^{'}=\frac 2\tau p$. Note that in the 
following figures the apex of the quantum variable $p^{'}$ is omitted.

In fig.8 the first moment of $\hat p$ and the Husimi function of the 
momentum distributions 
at different times are shown for $M=1, N=128$ and $k=4.42/\tau$ 
($\tau\approx 0.098$ and $k=45.022$). 
For the chosen value of $\tau$, the stable periodic orbits of the classical 
map (\ref{smapsim}) correspond to orbits $(\pi , \pm 15.6)$ 
and $(0,\pm 48.3)$ in the quantum phase-space.
The chosen initial state is a coherent state (\ref{inst}) centered in 
($\vartheta _0 =0, 
p^{'}_0 =\frac 2\tau *0.766\approx 15.605$).
Periodic tunnelling oscillations between the two symmetric islands 
centered in $\pm p^{'}_0$ take place.

The tunnelling oscillations are 
not suppressed even for long times. The calculation of $\langle p(t)\rangle$ 
in fig.8(d) is carried on for $10^6$ modulation periods.

Note that at $t=8400\tau \approx 824.668$, 
when $\langle p(t)\rangle$ assumes a maximum value, the quantum state is 
mainly localized 
in the island to the upper bound of the torus, centered approximately 
at $p^{'}_0 \approx\pm 48.3098$ (see fig.8(c)). 
As already said, the classical evolution of 
ensembles of points located in the islands 
centered in $ p_0 \approx\pm 2.4(p^{'}_0 \approx\pm 48.3)$ is independent from 
the evolution of points inside the islands
centered in $ p_0 \approx\pm 0.8(p^{'}_0 \approx\pm 15.6)$, 
in one of which the initial wave packet is 
localized.
\begin{figure}
\centerline{\epsfxsize=7cm \epsfbox{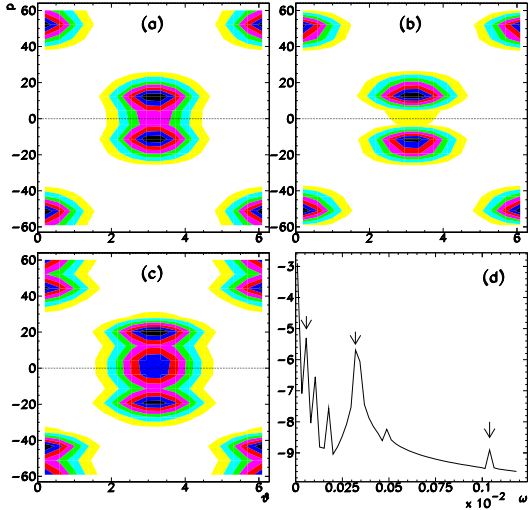}}
\caption{\small (a),(b),(c) The Husimi distribution of three eigenfunctions of 
the Floquet operator.
 (d) A section of the Fourier spectrum of $\langle p(t)\rangle$.}
\label{fig10}
\end{figure}

The quantum evolution instead couples structures that are independent in the 
classical system. In fact some eigenfunctions of the Floquet 
operator, involved 
in the dynamics, 
have high probability 
in both the pairs of islands (see fig.10 (a),(b),(c)).

\begin{figure}
\centerline{\epsfxsize=7cm \epsfbox{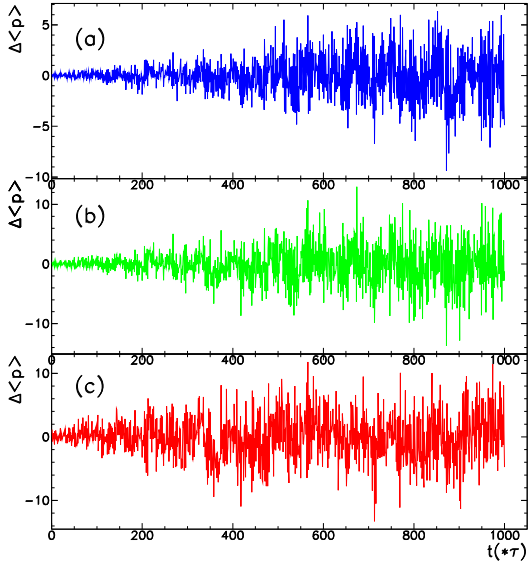}}
\caption{\small Suppression of periodical oscillations of the 
first  
moment $\langle p(t)\rangle$ in the presence of 
nonlinear perturbation ((a)
$u=5$, (b) $u=10$, 
(c) $u=20$). The difference 
$\Delta \langle p(t)\rangle$ is plotted. The values of the parameters  
are $\tau=4\pi /128$, $k=4.42/\tau , N=128, L=2000$.}
\label{fig11}
\end{figure}

In fig.9(a) a portion of the spectrum of the Floquet eigenphases versus $k$ is 
shown. 
An avoided crossing can be seen, close to the value of $k$ used 
in the calculations of this section, i.e. $k=\frac {2K}\tau\approx 45.022$.  
The corresponding
 eigenfunctions were selected by having a support in the region 
$|n|\leq 
25$ greater than $0.4$. The region $|n|\leq 25$ corresponds to the classical 
portion of the phase space $|p|=|\frac \tau 2 n|\leq 1.227$, 
in which the stability 
islands centered in $p_0 \approx \pm 0.7656$ lie.
Owing to the invariance properties of the system, the 
eigenfunctions localized in regular regions of the classical phase space 
occur in doublets of states with opposite symmetries and nearly degenerate 
eigenvalues; wave packets 
localized in the symmetric islands of stability are formed by symmetric 
or antisymmetric combinations of such doublets. In fig.9 (c),(d) an example 
of a quasi-degenerate doublet for $k=45.02$ is shown; the Husimi 
function of one state of this doublet is plotted in fig.10(b).

The dynamical tunnelling can be explained 
by a three states model \cite{LGrigW}.
In fig.9(b)-(l) the three eigenfunctions involved in the tunnelling 
process for three different values of $k$ are shown. Fixing a value of $k$, the
three states that take part in the dynamical tunnelling are a doublet of quasi-
degenerate states with opposite symmetry, 
localized in regular islands, and 
a third state localized 
in the chaotic region outside the stability islands ((b) for $k=45.02$,(e) for 
$k=45.05$,(l) for $k=45.08$).
It is the chaotic state that enhance dynamic tunnelling 
between stability islands.

By varying $k$, the chaotic eigenstate (fig.9(b) for $k=45.02$)
mixes itself with the state of the doublet sharing the same symmetry (fig.9(d) 
for $k=45.02$) until a complete exchange 
between the two states happens (fig.9(h) 
and fig.9(l) for $k=45.08$)\cite{MM,LGrigW}.
 
Beside the triplet of states shown in fig.9, 
there are other Floquet eigenstates that contribute to the 
behaviour of the first moment (\ref{pmed}) at fixed $k$; 
the separations between 
their eigenvalues correspond to the 
frequencies found by the decomposition of $\langle p(t)\rangle$ in Fourier 
components.

The frequencies mostly contributing to the motion are revealed by  
peaks of the power spectrum of $\langle p(t)\rangle$, 
represented in fig.10(d) for the parameter $k\approx 45.022$.
The spectrum has been calculated with a resolution of $\Delta \omega =2\pi
/2^{18}\approx 0.000024$. The three frequencies marked by arrows are
$\omega _1 = 0.000057\pm 0.000012,
\ \ \omega _2 = 0.000321\pm 0.000012,\ \ \omega _3 = 0.001040
\pm 0.000012$; 
$\omega _1$, $\omega _2$ and $\omega _3$ correspond to the difference between 
eigenphases
of three doublets of quasi-degenerate eigenfunctions: $\omega _1=\phi _{a}
-\phi _{a'}=5.501869-5.502808$, $\omega _2=\phi _{b'}
-\phi _{b'}=2.449073-2.448753$ and $\omega _3=\phi _{c'}
-\phi _{c'}=4.733879-4.732833$, where $\phi _{a},\ \phi _{b},\ \phi _{c}$ 
are the phases of the eigenfunctions 
plotted in fig.10(a),10(b),10(c) respectively.
These eigenfunctions have a support $\bar p$ 
in the region $|n|\leq 25$ greater than 0.45 and an overlap probability 
with the initial state $P_j$ greater than 0.01 ($P_j P_i > 0.0001$). 

Also in this case we may consider the effect of 
a nonlinear perturbation $u|\psi |^2$ in the Hamiltonian 
operator. The effect of this nonlinear perturbation on the quantum dynamics of 
the Kicked Rotator has already been studied in ref.\cite{cas,dima} but only in 
non resonant cases.

\begin{figure}
\centerline{\epsfxsize=7cm \epsfbox{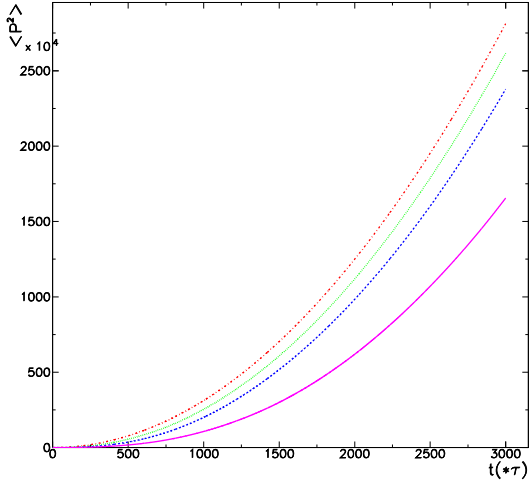}}
\caption{\small Second moment vs time, measured in numbers of periods for
different values of the nonlinearity: starting from above, $u=0$ 
dashed-dotted line, $u=5$ dotted line,
$u=10$ dashed line, $u=20$ full line. 
The values of the parameters  
are $\tau=4\pi $, $k=2.5 , N=16384, L=2000$}
\label{fig12}
\end{figure}

Since the perturbation depends 
continuously on time through the wave 
function $\psi$ of the system, the time evolution operator  
$\hat R$ between the kicks is approximated as 
a product of evolution operators on small intervals of time $\tau /L$ ($L$ is 
the number of small steps):
\begin{equation}
\label{evstatenl}
\hat R =\prod _{i=1}^{L} e^{-i\frac \tau {4L}n^2}
e^{-i\frac i2\frac \tau L |\psi _n |^2}
\end{equation}

The effect of a nonlinear perturbation is to break the symmetries of 
the system thus destroying tunnelling. 
In fig.11 the difference between the first moment of the unperturbed 
system ($u=0$) and 
that of the perturbed one $\Delta \langle p(t)\rangle 
=\langle p(t)\rangle_u -
\langle p(t)\rangle_0$ 
is plotted for $u=5,10,20$.

We have also analyzed the effect of the nonlinear perturbation on the 
second moment. 
In analogy with what has been found for the kicked harmonic oscillator 
under the resonance condition\cite{roblau}, in the presence of the nonlinear 
perturbation the growth of the second  moment is slower than in 
the kicked rotator, even if it remains ballistic 
($\langle p^2 (t) \rangle=c(\beta ) t^2$). 
In fig.12 the second moment vs time is shown: the four curves correspond 
to $u=0,5,10,20$ starting from above. This persistence of resonant behaviour 
once the nonlinear term is switched on (at least on our observation time scale)
is somehow surprising, and we hope further theoretical analysis will 
reveal its significance (i.e. if it not suppressed on longer time scales).

\subsection {Generic case.}
\begin{figure}
\centerline{\epsfxsize=7cm \epsfbox{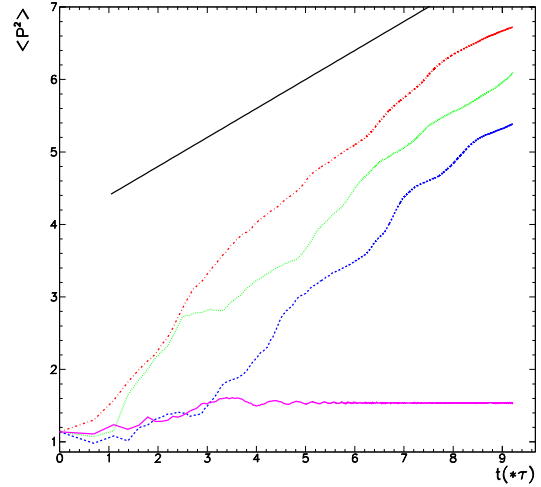}}
\caption{\small Suppression of dynamical localization of the 
time averaged second 
moment $P_{int}^2(t)$ in the presence of 
nonlinear perturbation (starting from below, 
$u=0$, full line, 
$u=5$, dashed line, $u=10$ dotted line, 
$u=20$,
dashed-dotted line). The parameters values are $\tau=1$, $k=2.5, N=128, 
L=2000$.}
\label{fig13}
\end{figure}
We consider the kicked rotator for $\tau =1$ and analyze the effect of the 
nonlinear perturbation on the first and second moment.
For this value 
of $\tau$ the system is far from the classical limit and displays quantum 
localization.
\begin{figure}
\centerline{\epsfxsize=7cm \epsfbox{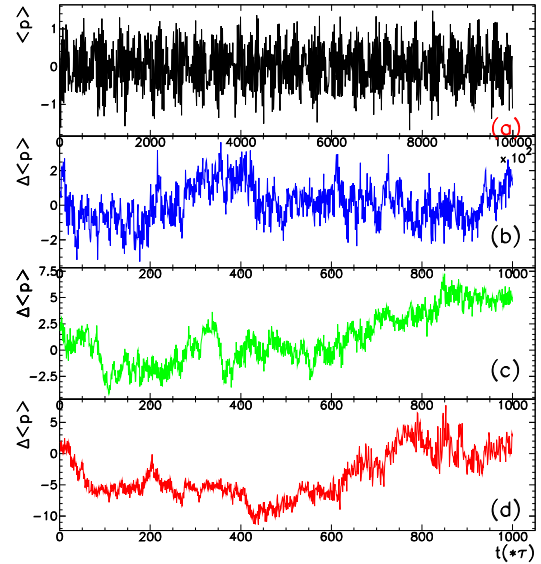}}
\caption{\small (a) Tunnelling oscillations in the non resonant Kicked Rotator;
(b),(c),(d) suppression of dynamical tunneling of the 
first moment $\langle p(t)\rangle$ in the presence of 
nonlinear perturbation ((b)
$u=5$, (c)$u=10$, 
(d)$u=20$). The difference 
$\Delta \langle p(t)\rangle$ is plotted. The 
parameters values 
are $\tau=1$, $k=4.42, N=128, L=2000$.}
\label{fig14}
\end{figure}
\begin{figure}
\centerline{\epsfxsize=7cm \epsfbox{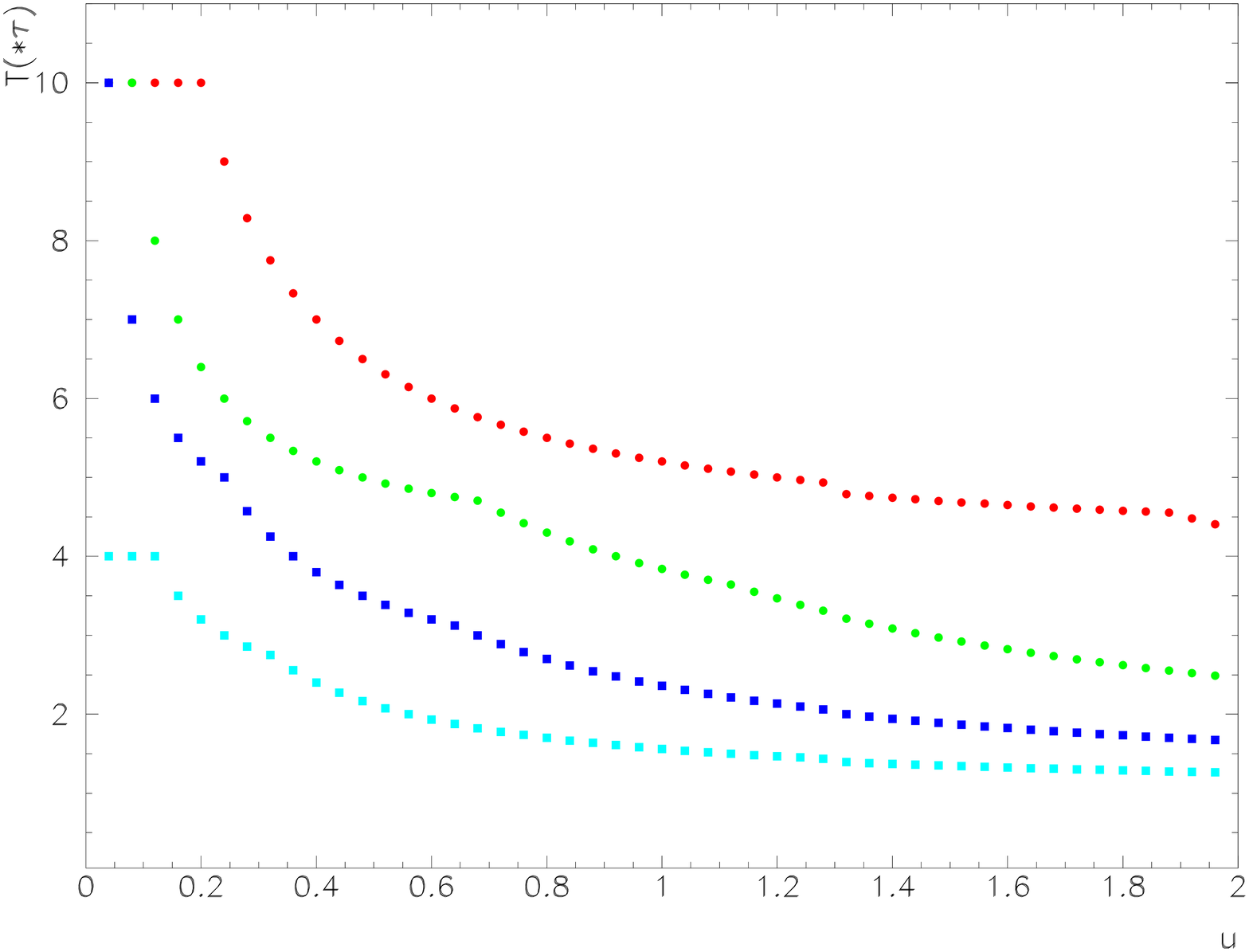}}
\caption{\small Time $T$ at which 
$|\Delta \langle p(t)\rangle |$ becomes 
greater than $\Delta$ versus the nonlinear parameter $u$. 
Each curve correspond to a different values of $\Delta$: 
$\Delta=0.4,0.2,0.1,0.05$ starting from above.}
\label{fig15}
\end{figure}

In accordance with \cite{dima}, 
the nonlinear perturbation suppresses localization 
and give rise to an anomalous diffusion with 
an exponent approximately equal to $2/5$.
In fig.13 a bilogarithmic 
plot of the time-integrated second moment 
$P_{int}^2 (t)=\frac 1t\sum _{t'=0}^{t-1}\langle p^2 (t')\rangle$ 
for $k=2.5$ and for 
different values of $u$ is shown; the 
straight line in the figure has a slope of $2/5$.

Tunnelling oscillations of $\langle p(t)\rangle$ for the Kicked Rotator are 
plotted in fig.14(a).
In Fig.14 (b),(c),(d), we show the suppression of tunnelling 
oscillations of the 
first moment for different values of 
the nonlinear parameter (b)$u=5$, (c)$u=10$, (d) $u=20$. 
The differences $\Delta \langle p(t)\rangle =\langle p(t)\rangle_u 
-\langle p(t)\rangle_0$ are plotted.

In Fig.15 the time $T$ at which $|\Delta \langle p(t)\rangle|$ becomes greater 
that $\Delta$ 
versus $u$ is plotted. Each curve corresponds to a different values of 
$\Delta$. Note that the evolution of $\langle p(t)\rangle$ in the kicked rotator 
is calculated 
at time intervals equals to $\tau$, therefore 
the estimate of the time $T$ is less
precise than that calculated for the driven pendulum.

\section {Conclusions.}
We have analyzed different physical systems for which dynamical tunneling 
occurs: in particular the driven pendulum and the kicked rotator. These 
systems are relevant for experimental settings recently realized with cold 
atoms. We have stressed several new features: from the role of how initially 
quasimomentum states are assembled, to the effect of Gross-Pitaevskii 
nonlinearities, in particular revealing subtle features happening in the 
resonant kicked rotator case.

{\bf Acknowledgments.}

This work was partially supported by EU contract QTRANS network (Quantum 
Transport on an Atomic Scale) and INFM PA project (Weak Chaos: theory 
and applications).

\begin{table}[h]
\begin{tabular}{||c|c|c||}
 $\Delta$ & $B$ & $A$ \\
\hline
0.005 & 0.0719$\pm$0.0012 & 0.1547$\pm$0.0096 \\
0.01 & 0.0836$\pm$0.0012 & 0.1649$\pm$0.0084 \\
0.02 & 0.0979$\pm$0.0011 & 0.1664$\pm$0.0072  \\
0.05 & 0.1219$\pm$0.0011 & 0.1908$\pm$0.0059 \\
0.1 & 0.1456$\pm$0.0010 &  0.2031$\pm$0.0050 \\
0.2 & 0.1766$\pm$0.0012 & 0.2193$\pm$0.0042 \\
\end{tabular}
\caption{Fitting parameters of $T$.}
\end{table}


\end{multicols}
\end{document}